# Experimental data management platform for data-driven investigation of combinatorial alloy thin films

Jaeho Song, Haechan Jo, and Dongwoo Lee*

School of Mechanical Engineering, Sungkyunkwan University, Suwon, Gyeonggi-do, South Korea

*Corresponding author, E- mail address: dongwoolee@skku.edu

**ABSTRACT**

Experimental materials data are heterogeneous and include a variety of metadata for processing and characterization conditions, making the implementation of data-driven approaches for developing novel materials difficult. In this paper, we introduce the Thin-Film Alloy Database (TFADB), a materials data management platform, designed for combinatorially investigated thin-film alloys through various experimental tools. Using TFADB, researchers can readily upload, edit, and retrieve multidimensional experimental alloy data, such as composition, thickness, X-ray Diffraction, electrical resistivity, nanoindentation, and image data. Furthermore, composition-dependent properties from the database can easily be managed in a format adequate to be preprocessed for machine learning analyses. High flexibility of the software allows management of new types of materials data that can be potentially acquired from new combinatorial experiments.

## 1. INTRODUCTION

Search for new alloys with an enhanced set of properties can be benefited from having diverse and large property datasets and advanced analysis tools such as machine learning [1-12]. The advent of high-throughput computation and experimentation has accelerated the exploration of composition spaces, especially for multicomponent systems, such as metallic glasses [13-18], High-entropy alloys [19-24], and magnetocaloric materials [25-27]. For these alloys, the unexplored composition region is immense, necessitating the construction of large materials datasets. Computational materials databases based on density function theory (DFT), such as AFLOW [10, 28], Materials Project [29], and OQMD [30] provide materials data of millions of different crystals, such as structures, enthalpies, electronic structures, etc. These databases have been successfully used for data-driven approaches to predict the material properties of crystals [31-33]. When it comes to multi-component alloys, however, limited information can be acquired from the simulation-based materials databases. This is mainly because many important properties of the alloys, such as strength, and electrical resistivity, are strongly dependent on the structural defects, whose features are not readily available in the DFT-based databases.

Combinatorial experiments [3, 34-40] allow facile acquisition of the property data of multi-component alloys. Thin-film alloys with composition spreads can readily be fabricated through physical vapor deposition (PVD) processes, such as evaporation [41, 42], electrodeposition [43, 44], and magnetron sputtering [34, 45-48]. Scanning high-throughput measurements on the combinatorial films can effectively produce composition-dependent materials data, such as microstructure, micrographs, as well as mechanical, electrical, thermal, and optical properties [34, 39, 48-53]. Materials data from the combinatorial experiments can be useful not only for designing advanced small-scale materials (e.g., interconnects and metal MEMS materials, etc. [52]) but also for predicting the properties of bulk materials [53].

Experimental materials data, however, are rather heterogeneous, multifaceted, and contain a variety of processing and characterization parameters (metadata), making the construction of the databases challenging.[54] Nevertheless, previous work demonstrated that a wide variety of metadata and data generated from combinatorial and high-throughput experiments can be successfully managed [55-59]. Soedarmadji et al. [55] demonstrated systematic management of millions of thin-film data, including structural, optical, and electrochemical properties as well as metadata from various characterization experiments. Banko, L. et al. [58] utilized a commercial document management system to manage the metadata from PVD processes and diverse characterization data, such as chemical, optical, structural, and magnetic properties.

The computational and experimental materials databases introduced above have achieved important advances in research data management (RDM) to study various classes of materials, although implementation of RDM and utilization of data-driven alloy investigation are still challenging for the research groups with a lack of related experiences. Here, we introduce the Thin-Film Alloy Database (TFADB), a RDM platform that is designed specifically for thin-film alloys investigated by combinatorial experiments. The database manages the metadata and the property data obtained from high-throughput experiments. TFADB is a ready-to-use platform that is designed to be easily implemented by the research groups to construct their own alloy databases, quickly overcoming the obstacles to the employment of customized RDMs. The database is also readily managed, edited, and re-configured through the user-friendly GUI (graphical user interface).

The open version of TFADB (see *data availability*) can manage the composition, electrical resistance, thickness, X-ray diffraction, nanoindentation, and sample image data (optical microscope, scanning electron microscope, and transition electron microscope images) of combinatorially investigated thin films. New types of data from other experimental tools can also be uploaded and managed through TFADB. Furthermore, TFADB can output the experimental datasets in a machine-readable format, facilitating the implementation of data-driven approaches to develop new alloys.

## 2. RESULTS AND DISCUSSION

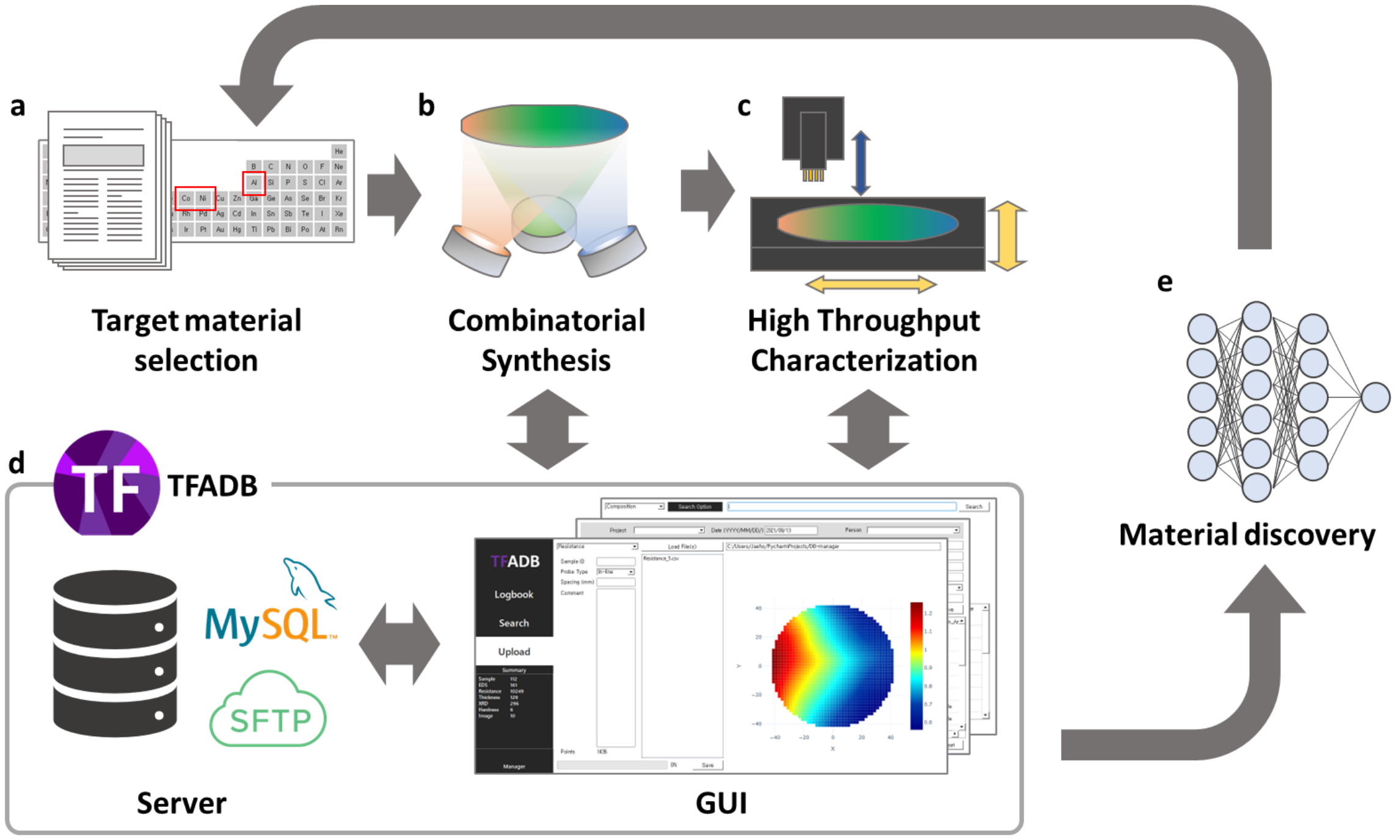

**Figure 1. Alloy development process using TFADB** (a) The target composition range for a combinatorial study is determined by the domain knowledge or the property prediction model that is based on previously deposited materials data available in TFADB. (b) Combinatorial synthesis of the target composition is carried out. Deposition histories available in TFADB can be used to determine the processing parameters of the new deposition. (c) Scanning property measurements (high-throughput experiments) measure position-dependent properties. Since the composition of the combinatorial specimens varies with respect to the position, position-dependent properties can be converted to composition-dependent properties. (d) The data are uploaded to TFADB. The uploaded data can be managed, displayed, and downloaded by multiple users. (e) Systematic data management of TFADB allows implementation of data-driven approaches for new alloy design.

Fig. 1 schematically illustrates the data-driven novel alloy design procedure that utilizes TFADB. The alloy candidates with target compositions and properties for the synthesis are selected based on the domain knowledge of the researcher or the property prediction model that was built using previously deposited materials data in TFADB (Fig. 1a, e). Combinatorial thin film alloys with the target composition range are then synthesized by reviewing the processing parameters (metadata) of similar alloy systems that may be available in TFADB (Fig. 1b). Combinatorial magnetron sputtering without a substrate rotation is one of the common choices to produce combinatorial thin films: films with different compositions at different positions on a substrate [3, 34, 40, 45-48, 52, 60]. Typically, hundreds of alloys with different compositions can be synthesized from a single deposition. Processing parameters of the PVD process, such as Ar pressure, substrate temperature, sputter gun power, and gun angle affect the composition ranges and properties of the combinatorial specimens. Therefore, the metadata, recipe of the synthesis process and the information related to the thin film sample such as project name, date, and researcher, are systematically managed in TFADB (Fig.1d).

The combinatorially synthesized films are then subjected to scanning property measurements and position-dependent properties are acquired and uploaded to TFADB (Fig. 1c). Different kinds of materials data can be acquired from conventional experiments that support the scanning measurement modes, such as XRD (X-ray diffraction), SEM (scanning electron microscope), and nanoindentation. Custom-designed combinatorial techniques can also be used for thermal [39, 61-64], electrical [3, 34, 40, 52, 60], and magnetic [49, 65, 66] properties. Acquisition of the materials data from different sets of experiments and tools results in heterogeneity in the data format. Therefore, the establishment of the standard format of the characterization data is required as will be discussed later. The users of TFADB are encouraged to upload the metadata and property data of combinatorial alloy films from unsuccessful experiments, such as poor materials properties (e.g, low hardness) or the processing parameters of a problematic synthesis (e.g, peeling-off of thin films). This will not only help future researchers find adequate processing parameters for new alloy films but also enriches the diversity of the materials data which are beneficial to build more reliable machine learning models [67].

In the last step of the data-driven procedure for the alloy development using TFADB, the new property data can be compared with the result from the prediction model (Fig. 1e). Good agreement between the predicted and measured properties indicates new materials with desired property is developed. If the model accuracy is found to be low, new machine learning models can be constructed by using the newly acquired data. Iterative processes of new combinatorial data acquisition and prediction model improvement would be needed to design alloys with better properties.

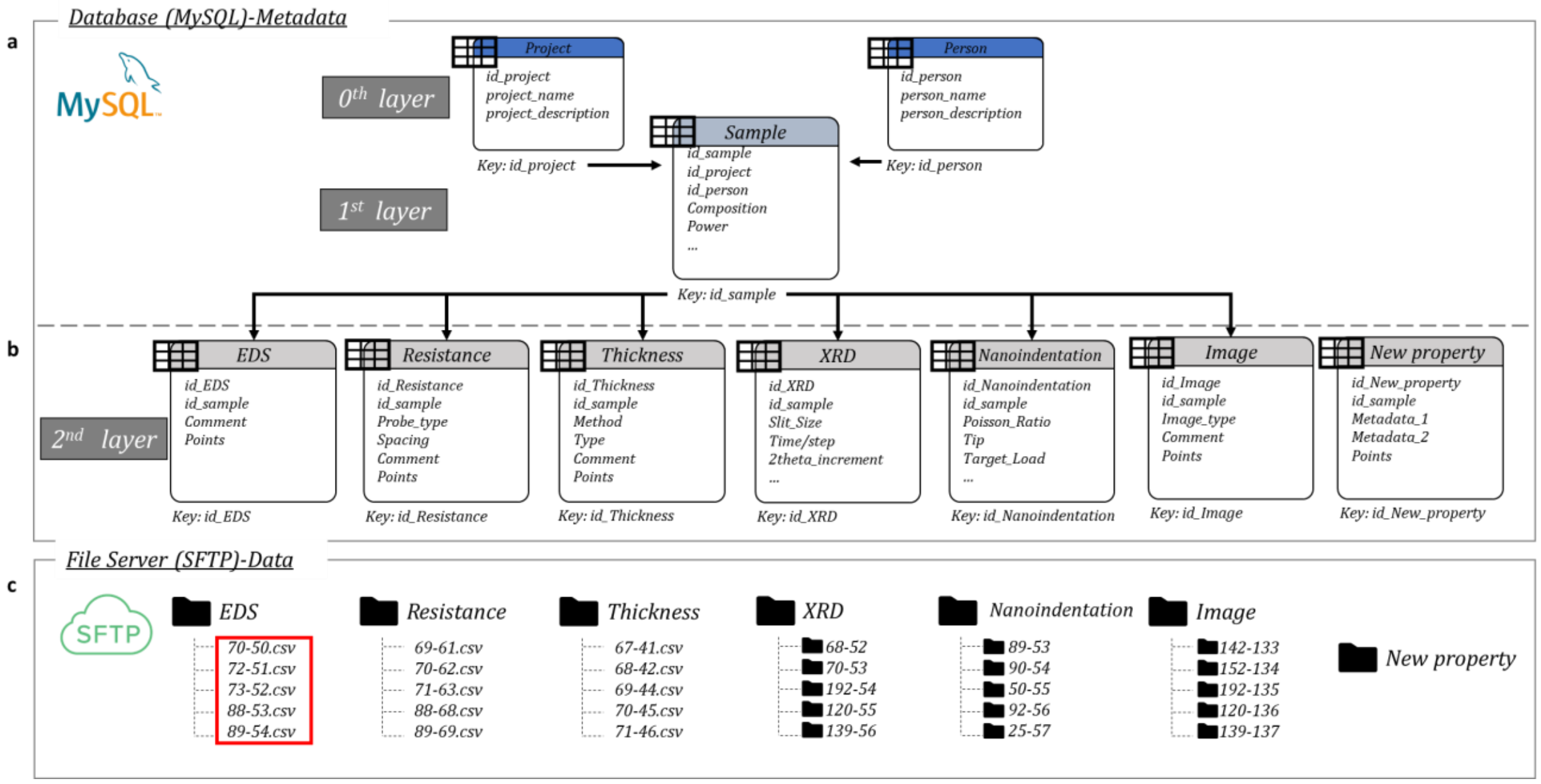

**Figure 2. Hierarchical structure of TFADB** (a) The metadata of combinatorial specimen fabrication constitute the 0$^{th}$ and the 1$^{st}$ layers of the MySQL database. (b) The metadata of the characterizations are managed in the 2$^{nd}$ layer of the MySQL. (c) The measured properties of the combinatorial specimens are uploaded to the SFTP file server. Each metadata or data is assigned with an *id key* allowing tracking of data lineage. The one-dimensional property data with respect to a position on a wafer are stored in a .csv file (e.g. EDS, resistance, thickness data), while the multi-dimensional data (e.g. XRD, nanoindentation, image data) are managed by saving them in directories.

The combinatorial synthesis and high-throughput characterizations mentioned earlier produce the metadata of processing conditions and characterization tools as well as the property and image data. The hierarchical structure of TFADB illustrated in Fig. 2 make these different types of data linked together and traceable. Different sets of property data and metadata in TFADB are managed through two different servers of MySQL and SFTP (Fig. 2). The metadata are saved in the MySQL database as hierarchical tables (Fig. 2a-b). Specifically, the metadata such as researcher and project information are saved on the $0^{th}$ layer of the DB structure as *Person* and *Project* tables (Fig. 2a). *Sample* table collects the processing parameters of the combinatorial samples (*e.g.*, target composition, chamber pressure, deposition power, Ar gas pressure, deposition temperature, annealing temperature, etc.). *Sample* table is practically used as the first layer of the database hierarchy, and each sample metadata is assigned with the primary key, "*id_sample*" for tracking data lineage. *Sample* table contains the project and person information by inheriting the "*id_project*" and "*id_person*" columns. The characterization metadata tables constitute the $2^{nd}$ layer (Fig. 2b) and collect the information of measurement conditions of the characterization process. These tables inherit "*id_sample*" from the sample table.

The characterization data that are obtained from various experimental tools are saved in the local directory of the SFTP server, whether as files or folders, depending on the dimension of the data (see Fig. 3). The characterization data files in the SFTP server are managed by labeling the name of the file/folder identical to the primary key id of the sample and characterization process metadata in a format of "*id_sample-id_property*" (as shown in the red box in Fig.2 c for example). The labeled *id* makes the data findable through the primary keys of the metadata, enabling the association between different characterization data for a same specimen. Using the data management strategy illustrated above, addition of a new characterization process can be readily done.

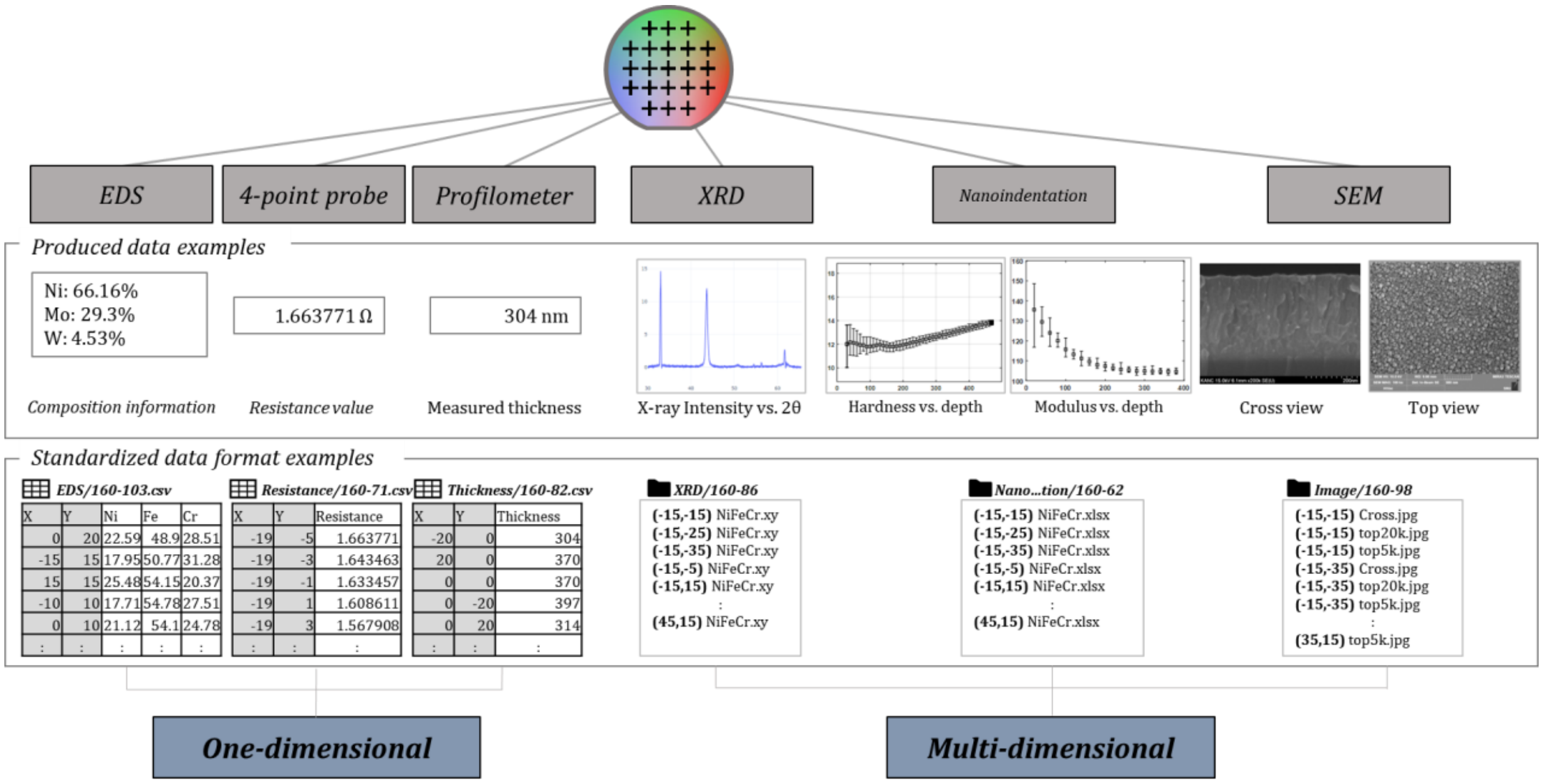

**Figure 3. Categorization of materials data in TFADB** The position-dependent characterization data from high-throughput experiments are categorized into one-dimensional (e.g. composition, resistance, thickness) and multi-dimensional (e.g. XRD, nanoindentation, SEM images, etc.) data.

The high-throughput characterization processes produce position-dependent materials information. Since combinatorial alloy films have compositional gradients across the in-plane direction of the wafer, composition-dependent properties can be acquired once the position-composition relation is found. Different characterization methods, however, have different dimensions in the materials data, requiring different strategies for managing the data (Fig. 3). Materials data such as composition, thickness, and electrical resistance at a point of the specimen are categorized as one-dimensional data. For each position, unique materials information represented by texts and/or numbers are determined. The characterization data of the one-dimensional data is stored in tables with the positional information. Materials data acquired from XRD, nanoindentation, and SEM at a point have two or higher dimensions. For these types of data, the materials information for each position are saved in a file named "*(x, y position)filename*". With this method, multiple data from the same technique at a point (e.g., in-plane and cross-sectional SEM images acquired from the same position or composition) can be saved and managed.

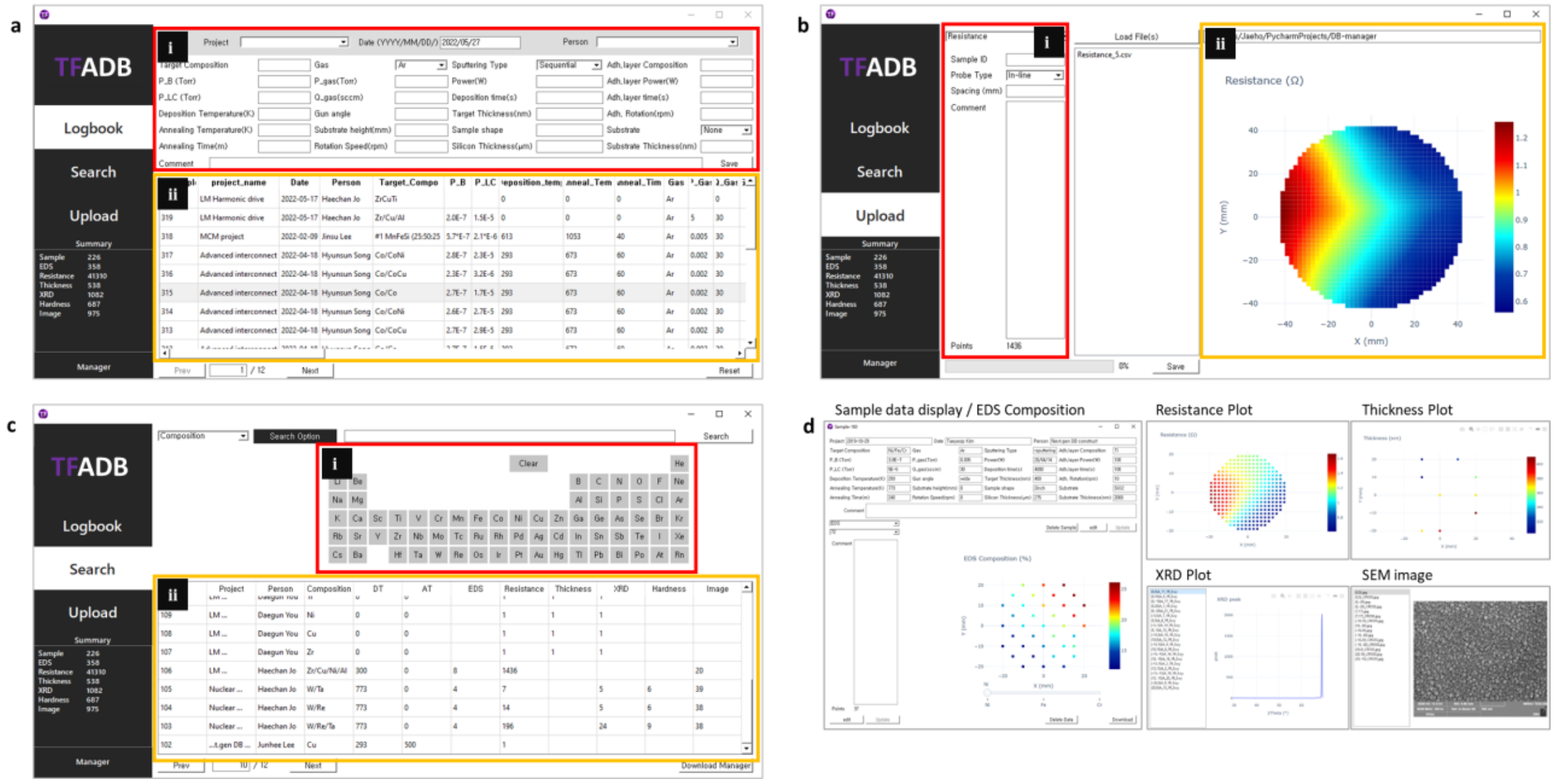

**Figure 4. Graphic user interface of TFADB** (a) *Logbook* tab allows the users to register a new specimen with the information of the deposition conditions used (a-i). The history of the logbook information can be accessed in the window a-ii. (b) *Upload* tab is used to upload the thin film property data along with the metadata of measurement conditions (b-i). The uploaded materials data can be displayed as shown in the window b-ii. (c) Through *Search* tab, users can search materials data of interest in TFADB by typing the keywords (compositional information, project, person, and sample id, etc.) or selecting the elements in the periodic table

(c-i). Summary of the retrieved information can be seen in the window c-ii, and all the related information can be downloaded. (d) The loaded materials information, such as composition, resistance, and thickness maps as well as XRD results and images can also be displayed in the GUI.

Through the graphic user interface of TFADB, the users can upload, search, and download the materials property data and the metadata (Fig. 4). *Logbook* tab in the GUI is used to register a new combinatorial specimen and save the information of the processing parameters, such as the sputter targets used, power of each gun, deposition time, base pressure, working gas and pressure, deposition temperature, substrate type, researcher name, and deposition date (Fig. 4a-i). The processing information of previously deposited films can be retrieved (Fig. 4a-ii). This electronic logbook is helpful when a new researcher determines the processing conditions for new target compositions. *Upload* tab of the GUI allows the users to upload the characterization data and the related metadata to TFADB (Fig. 4b). To upload the data, the users need to insert the *sample id* and the metadata of the characterization processes and load the characterization data (Fig. 4b-i). The uploaded characterization data can be viewed in the window as exemplified in Fig. 4b-ii. Through *Search* tab of the GUI (Fig. 4c), the users can find, download, update, and delete the uploaded characterization data and metadata of the selected samples from the database. Combinatorial specimens of interest can be found by inserting the keywords of composition, sample id, project name, and researcher name or by selecting the constituent elements of the interested alloy from the periodic table (Fig. 4c-i). When the selected information is available in the database, a summary of the list can be viewed in the window in Fig. 4c-ii. All the selected data are simultaneously downloadable in CSV format using the download manager available in *Search* window. Various characterization data of selected samples can also be displayed through the GUI (Fig. 4d). These features of the GUI minimize the effort of the researchers for data processing and feature preparation for the machine learning implementation and meet the FAIR (Findability, Accessibility, Interoperability, and Reusability) data principles [54, 68].

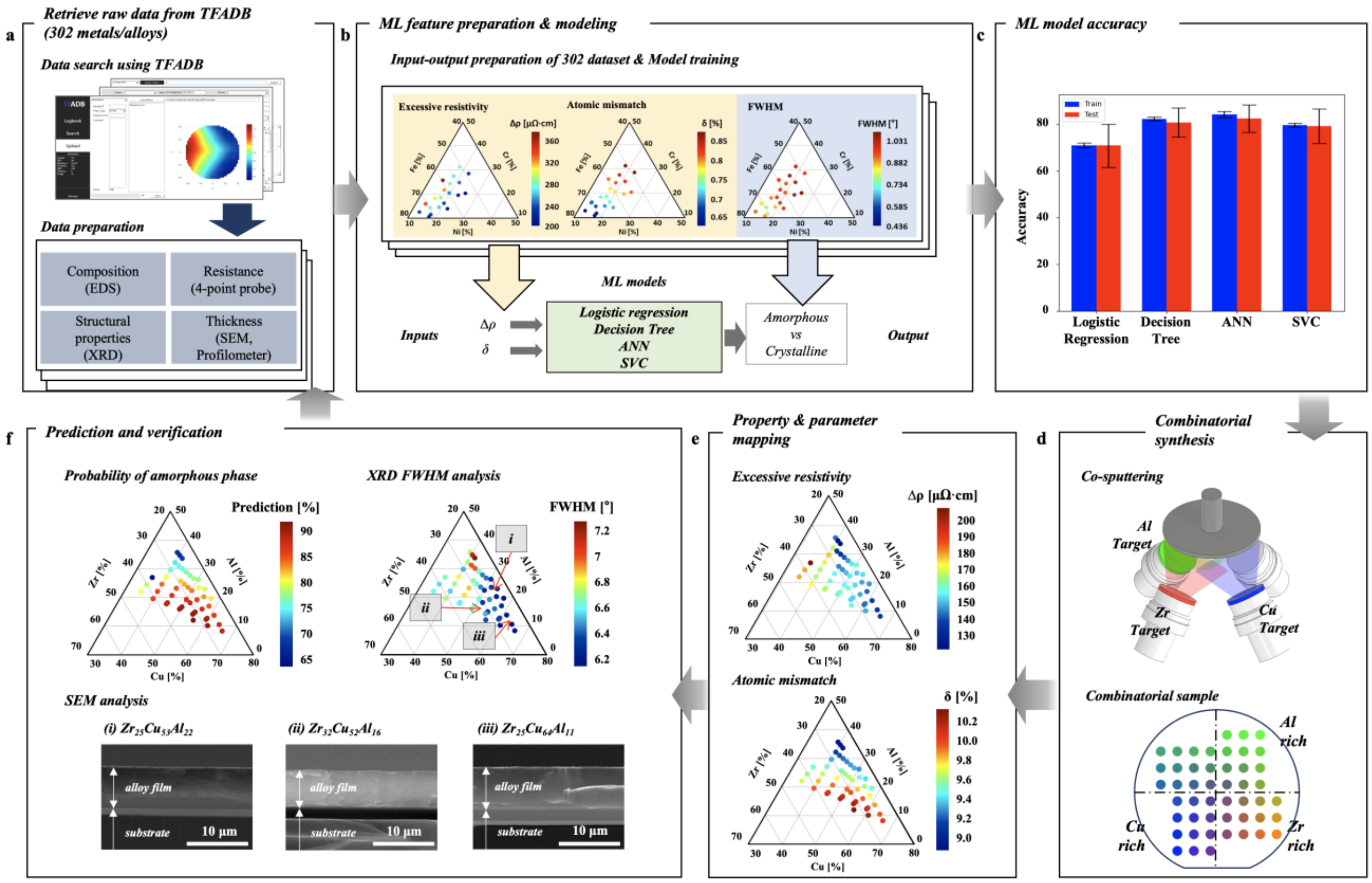

**Figure 5. Example of machine learning implementation using TFADB to classify amorphous/crystalline phases of alloys** (a) Composition, electrical resistance, thickness, and XRD data of 302 thin film alloys were retrieved from TFADB. (b) The retrieved data were processed and used as training datasets for the machine learning models to classify amorphous/crystalline phases of the alloys; excessive resistivity (Δρ), atomic mismatch (δ), and the FWHM of the XRD peaks respect to compositions were acquired. The prepared features were fed into the machine learning classifiers based on logistic regression, decision tree, ANN, and SVC. (c) The prediction accuracies of the models were evaluated with 10-fold cross-validation. The test accuracy of ANN (82.45%) was the highest among the models. (d) Combinatorial synthesis of a new thin film alloy system ($Cu_{34.9\text{-}66.3}Zr_{23.1\text{-}43.1}Al_{8.2\text{-}35.7}$) was carried out, followed by property mapping. The new dataset was used to test the ANN model built. (e) High-throughput characterizations of Cu-Zr-Al sample were performed, and features of the machine learning model (Δρ and δ) were prepared to predict the phase map of the combinatorial thin film specimen. (f) The probability map of amorphous phase is compared with the measured FWHM map for the Cu-Zr-Al specimen. Both results indicate that all the compositions are in the amorphous state. The amorphous phase formation of selected compositions was double confirmed by cross-sectional SEM images in (f-*i*~*iii*).

In Fig. 5, we demonstrate how the materials data available in TFADB can be employed and processed for machine learning implementation. We built machine learning models to classify amorphous or crystalline phases of the alloys by using the excessive resistivity (Δρ) and determining the atomic mismatch (δ) as a function of composition. These models would allow efficient phase identification for unknown alloy systems without utilizing slow or expensive experimental tools [3]. To train the classifiers, we utilized materials data of 302 different metals and alloys, including pure metals (Zi, Ti, Ni, Al, Cu), as-deposited, and annealed combinatorial systems (Cu-Zr, Cu-Ti, Ni-Zr, Ni-Ti, Cu-Zr-Ti, Ni-Fe-Cr), and as-deposited combinatorial systems (Al-Cu-Zr, Al-Cr-Ti) that have been deposited to our own TFADB (Fig. 5a). The characterization data of EDS, thickness (*t*), electrical resistance (*R*) were used to determine composition-dependent Δρ using $\Delta\rho = \rho - \rho_{rom}$, where $\rho = FRt$ and $\rho_{rom} = \Sigma_{i=1}^{N} c_i \rho_{i,\,bulk}$ (Fig. 5b) [3]. Here, $F$, $c_i$, $\rho_{i,\,bulk}$, and *N* respectively are the correction factor [69, 70], composition and bulk resistivity of the constituent element *i*, and number of the constituent elements of the alloy. δ for a composition can be simply calculated using $\delta = \sqrt{\Sigma_{i=1}^{N} c_i \left(1 - \frac{r_i}{r_{avg}}\right)^2}$, where $r_{avg} = \Sigma_{i=1}^{N} c_i r_i$. Here, $r_i$ is the atomic radius of the element *i*. For the output, i.e., amorphous/crystalline phases, an alloy was labeled as an amorphous phase if the measured FWHM value from the XRD data is greater than 3.4° [71]. Other cases were labeled as crystalline (Fig. 5b).

The artificial neural network (ANN), logistic regression, decision tree, and support vector classifier (SVC) were considered as phase prediction models as detailed in *methods*. Each model was trained with the same set of features and outputs of the 302 alloys with the 10-fold cross-validation and the results showed that the ANN model have the highest train (84.2%) and test (82.5%) accuracies (Fig. 5c). A new combinatorial specimen with a composition range of $Cu_{34.9\text{-}66.3}Zr_{23.1\text{-}43.1}Al_{8.2\text{-}35.7}$ was fabricated to test the ANN model (Fig. 5d). The deposition parameters of the combinatorial synthesis of the Cu-Zr-Al system were determined by retrieving and employing the metadata of the processing conditions of the elemental targets of pure Cu, Zr, Al from TFADB. The synthesized combinatorial specimen was then subjected to composition, thickness, electrical resistance measurements, allowing determination of the features of Δρ and δ (Fig. 5e). The ANN model with the two features resulted in the probability map for amorphous phase of the Cu-Zr-Al system as shown in Fig. 5f. The model showed that all the compositions have the probability greater than 64.7 %, indicating that only the amorphous phase was expected to be formed regardless of the composition of the system. This prediction result was compared with the XRD and cross-sectional SEM images for selected compositions: all the compositions were indeed in amorphous state.

This example demonstrates that previously deposited data of thin film alloys can be retrieved and processed using TFADB for the data-driven analysis of combinatorial thin film alloys. Fabrication and characterizations of new targeted combinatorial alloys were also facilitated by retrieving the saved metadata of similar or related samples in TFADB. Continuous depositions of materials data using TFADB will enrich the quantity and diversity of the database, allowing effective implementation of data-driven approaches for alloy investigations with high reliability. TFADB is a ready-to-use and open-source software with high flexibility (see data availability). We believe that TFADB meets the growing need of the research groups for the adaptation of data-driven alloy investigations with ease.

## 3. CONCLUSIONS

A novel research data management platform, TFADB, to track lineage of the experimental data of alloys from combinatorial synthesis and high-throughput experiments has been developed. The hierarchical database structure to systematically manage heterogenous experimental data and the GUI to efficiently upload, retrieve, edit, and download the materials data and metadata are introduced. Examples of data management for the combinatorial sputtering process and the scanning measurements (composition, thickness, X-ray Diffraction, electrical resistance, nanoindentation, images) and implementation of data-driven approaches for alloy investigation have been discussed. The features of TFADB allow users to build more reliable prediction models by using previously deposited data and to readily determine synthesis parameters of unknown alloy systems. TFADB is a ready to use platform with high flexibility. We believe that TFADB can offer the opportunity to the research groups who seek the benefits of ever-growing materials data and to implement data-driven discovery of new alloy systems with ideal set of properties.

## 4. METHODS

### Development of TFADB platform and its servers

TFADB in this study was constructed with two Linux servers of an SQL and an SFTP file in a local NAS (Network Attached Storage) device. The SQL server consists of two data schemes, one having metadata of the sample synthesis and characterization processes, and the other with configurative settings of the database server. The SFTP file server consists of various characterization data acquired from high-throughput experiments. The GUI of the TFADB was constructed based on Python, with the PyQT5 library. The source code TFADB is open to the community through GitHub (*Data availability*).

### Combinatorial synthesis and high-throughput characterization of the Zr-Cu-Al thin film sample

The combinatorial alloy thin film of Zr-Cu-Al was magnetron-sputtered on the 4-inch wafer, under the base and Ar pressure of $2.0 \times 10^{-7}$ Torr and $5 \times 10^{-3}$ Torr, respectively. The deposition temperature and time were set to 293 K and 1,300 seconds, respectively. Elemental sputter targets of Zr, Cu, and Al were used for the depositions without substrate rotation to produce an in-plane composition spread over the substrate. A combinatorial film with a thickness ranging from 285 to 760 nm was produced, with a composition range of $Zr_{43.1\text{-}23.1}Cu_{66.3\text{-}34.9}Al_{8.2\text{-}35.7}$. The composition was evaluated by EDS (FESEM, JEOL JSM-7600F) and the thickness was measured by a profilometer (Bruker DXT-A). The electrical resistance map of the thin film was determined by using a custom-built high-throughput 4-point probe x-y-z stage system with a digital multimeter (RIGOL 3058e) and with probes aligned in-line, with 1 mm spacing. The XRD map of the combinatorial film was measured using a XRD (Bruker D8 advance eco) with the Bragg-Brentano geometry and using Cu-$K_\alpha$ radiation.

**Machine learning analyses**

The logistic regression, decision tree, and SVC models used in this study were built with the Scikit-learn library of Python. The hyperparameters of the models were set as default, except for the depth of 3 for the decision tree model to avoid overfitting. The ANN model built was based on the Keras library of Phyton. The hyperparameters were tuned based on the grid-search by varying the number of neurons from 1 to 100, the number of hidden layers from 1 to 2, and the learning rate from 1e-4 to 1e+0. As a result, the ANN model with 1 hidden layer, 50 neurons (ReLU), Sigmoid output, and a running rate of 1e-2 was chosen and trained.

**Acknowledgment**

This work was supported by Samsung Electronics Co., Ltd. (IO201211-08077-01), Samsung Future Technology Incubation Program (SRFC-MA1802-06), the National Research Foundation of Korea (NRF) (NRF- 2017R1E1A1A01078324, 2020M3D1A101609221, 2021R1A4A1029780), Fundamental Research Program of the Korea Institute of Material Science (PNK8270)

**Data availability**

The source code including the server and GUI, executable program, and manual for TFADB are available at https://github.com/jh-song-en/TFDMP.

**Author contributions**

J.S. and D.L. conceived and designed the TFADB RDMs, which was developed by J.S. H.J. carried out the combinatorial experiments for the thin film sample. J.S. wrote the paper and D.L. supervised the research and revised the manuscript. All authors discussed and commented on the manuscript.

**Competing Interests**

The Authors declare no Competing Financial or Non-Financial Interests